\documentclass[useAMS,usenatbib,onecolumn]{mnras}
\usepackage{amsmath,bm,amssymb}
\usepackage{dcolumn}
\usepackage[utf8]{inputenc}
\usepackage{graphics,graphicx}
\usepackage{float}
\usepackage{threeparttable}
\usepackage{url}
\usepackage{epstopdf}
\usepackage{soul}

\usepackage{booktabs}
\usepackage{multirow}

\usepackage{color}

\newcommand{\X}{$\tilde{X}\,^1\Sigma^+$}

\newcommand{\EVEREST}{{\sc EVEREST}}

\title[ExoMol line lists -- LI. LiOH]{ExoMol line lists -- LI. Molecular line list for lithium hydroxide (LiOH)}

\date{\today}

\author[A. Owens et al.]
{Alec Owens$^{1}$\thanks{The corresponding author: alec.owens.13@ucl.ac.uk},
Sam O.~M. Wright$^{1}$,
Yakiv Pavlenko$^{2}$, Alexander Mitrushchenkov$^{3}$, Jacek Koput$^{4}$, \newauthor{Sergei N. Yurchenko$^{1}$\thanks{The corresponding author: s.yurchenko@ucl.ac.uk} and Jonathan Tennyson$^{1}$\thanks{The corresponding author: j.tennyson@ucl.ac.uk}}\vspace*{4mm}\\
$^1$ Department of Physics and Astronomy, University College London, Gower Street, WC1E 6BT London, UK\\
$^2$ Main Astronomical Observatory, Academy of Sciences of the Ukraine, 27 Zabolotnoho, Kyiv 03143, Ukraine\\
$^3$ MSME, Universit\'{e} Gustave Eiffel, CNRS UMR 8208, Univ Paris Est Creteil, F-77474 Marne-la- Vallée, France \\
$^4$ Department of Chemistry, Adam Mickiewicz University, 61-614, Poznan, Poland}

\date{Accepted XXXX. Received XXXX; in original form XXXX}

\pagerange{\pageref{firstpage}--\pageref{lastpage}} \pubyear{2023}

\begin{document}

\label{firstpage}

\maketitle

\begin{abstract}
A new molecular line list for lithium hydroxide ($^{7}$Li$^{16}$O$^{1}$H) covering wavelengths $\lambda > 1$~$\mu$m (the 0\,--\,10\,000~cm$^{-1}$ range) is presented. The OYT7 line list contains over 331 million transitions between rotation-vibration energy levels with total angular momentum up to $J=95$ and is applicable for temperatures up to $T\approx 3500$~K. Line list calculations are based on a previously published, high-level \textit{ab initio} potential energy surface and a newly computed dipole moment surface of the ground \X\ electronic state. Lithium-containing molecules are important in a variety of stellar objects and there is potential for LiOH to be observed in the atmospheres of exoplanets. This work provides the first, comprehensive line list of LiOH and will facilitate its future molecular detection. The OYT7 line list along with the associated temperature- and pressure-dependent opacities can be downloaded from the ExoMol database at \href{http://www.exomol.com}{www.exomol.com} and the CDS astronomical database.
\end{abstract}

\begin{keywords}
molecular data – opacity – planets and satellites: atmospheres – stars: atmospheres – ISM: molecules.
\end{keywords}

\section{Introduction}
Ultracool dwarfs occupy the right-bottom corner of the Hertzsprung–Russell diagram. The rates of depletion of lithium (and deuterium) in their interiors are primarily determined by the mass of the star or substellar object. For stars $M > 85$~$M_J$, the burning of lithium, Li ($p,\alpha$) $^4$He, becomes efficient at early evolutionary stages preceding the main sequence at interior temperatures of $T\sim2.5$~MK~\citep{dant98}. Young, low-mass stars possess developed convective envelopes so lithium depletion in their central region occurs on relatively short time scales (several tens of millions of years), resulting in a weakening or complete disappearance of lithium lines in their spectra. The temperature in the interior of brown dwarfs with masses less than 60~$M_J$ is not high enough for lithium burning. Thus, in principle, their primordial lithium abundance should not change with time. This circumstance led to the idea of the lithium test~\citep{rebo92}, which essentially equates to searching for lithium absorption lines in the spectra of ultracool dwarfs as evidence of their substellar nature.

The lithium abundance in the atmosphere of young stellar objects is of special interest. In particular, the position of the line in the Hertzsprung–Russell diagram separating stars burning lithium from lower-mass objects with lithium still in their atmospheres can be used to derive independent age estimates for young open clusters less than 150 million years in age~\citep{dant98}. \citet{pavl95} showed that despite strong blending and formation of lithium-containing molecules, observation of Li resonance doublets is possible in M-dwarfs despite the strong presence of TiO and VO. Indeed, \citet{rebo96} observed Li resonance doublets in the spectrum of the first brown dwarf Teide 1 and rather high lithium abundance $\log N\mathrm{(Li)}=3.2\pm 0.2$~\citep{pavl97tc}. The lithium test was successfully used to prove the substellar nature of several late-type dwarfs of spectral class M~\citep{ruiz97,zapa02}.


In the case of late L- and T-dwarfs, the numerous molecular lithium-containing species bind most of the Li atoms~\citep{tsuj73}. Li I is not a dominant species, but lithium hydroxide (main isotopologue $^{7}$Li$^{16}$O$^{1}$H, which we refer to as LiOH from here on in) is one of the most abundant molecules, along with LiF and LiCl~\citep{pavl00}.
We note that line lists for LiF and LiCl were created by \citet{18BiBexx.LiF} as part of the MOLLIST project~\citep{MOLLIST} and are also available
from ExoMol~\citep{jt810}. For cooler T- and Y- dwarfs, the maximum of the spectral energy distribution shifts towards the infrared and fluxes in the optical spectral range become extremely weak. Moreover, due to the lower temperatures, lithium presumably exists in the form of molecules. Thus, the computation of spectral features formed by lithium-containing species such as LiOH is the only way to determine the lithium abundance in their atmospheres~\citep{ghar21}.

There is very limited knowledge of the gas-phase rotation-vibration (rovibrational) spectrum of lithium hydroxide. Pure rotational transitions in the ground vibrational state and excited $\nu_2$ bending vibrational states have been measured~\citep{94McTaKl.LiOH,04HiFrKl.LiOH}, confirming the linear equilibrium geometry of LiOH. \citet{96Gurvich.LiOH} reviewed the available data on thermodynamic and molecular properties of gas-phase LiOH, highlighting the lack of experimental studies. Theoretical calculations have been carried out using full-dimensional \textit{ab initio} potential energy surfaces (PESs)~\citep{89BuJeKa.LiOH,04HiFrKl.LiOH,13Koputx.LiOH} providing key insight into the rovibrational energy level structure, however, without essential information on the strength of transition intensities the chances of observing this molecule in astronomical environments is limited. To this end, we present a newly-computed molecular line list of $^{7}$Li$^{16}$O$^{1}$H covering infrared wavelengths $\lambda > 1$~$\mu$m (0\,--\,10\,000~cm$^{-1}$). The new line list, named OYT7, is available from the ExoMol database~\citep{jt810,jt631,jt528}, which is providing comprehensive molecular data to aid the atmospheric modeling of exoplanets and other hot astronomical bodies.
We mention that in the Solar System, lithium abundance is composed of two stable isotopes $^{7}$Li and $^{6}$Li with approximately 92.4\% and 7.6\% abundance. In this work we are only concerned with $^{7}$Li$^{16}$O$^{1}$H.

The paper is structured as follows: In Sec.~\ref{sec:methods} we describe the computational setup and theoretical spectroscopic model used to generate the LiOH line list. Results are discussed in Sec.~\ref{sec:results}, where we detail the structure and format of the line list along with generated opacities, analyse the temperature-dependent partition function, and simulate spectra of LiOH with a focus on potential detection in exoplanet atmospheres. Conclusions are offered in Sec.~\ref{sec:conc}.

\section{Methods}
\label{sec:methods}

The computational procedure for generating molecular line lists is well established~\citep{jt654,jt626} with a substantial number of line lists having been produced for the ExoMol database. Calculations require molecular PESs, dipole moment surfaces (DMSs), and a variational nuclear motion program to solve the Schr\"odinger equation to obtain rovibrational energy levels and all possible transition probabilities between them.

\subsection{Potential energy surface}

We utilise a previously published \textit{ab initio} PES of the \X\ electronic ground state of LiOH~\citep{13Koputx.LiOH}. The PES was computed using coupled cluster methods in conjunction with a large augmented correlation-consistent basis set (CCSD(T)/aug-cc-pCV6Z level of theory) and treated a range of higher-level additive energy corrections. These included core-valence electron correlation, higher-order coupled cluster terms beyond perturbative triples, scalar relativistic effects, and the diagonal Born-Oppenheimer correction. Computing the PES with a composite approach like this can be very accurate, predicting fundamental vibrational wavenumbers to within 1~cm$^{-1}$ accuracy on average. For $^7$LiOH, the fundamental vibrational wavenumbers are predicted to be $\nu_1=925.37$~cm$^{-1}$ for the Li--O stretching mode, $\nu_2=319.49$~cm$^{-1}$ for the Li--O--H bending mode, and $\nu_3=3833.14$~cm$^{-1}$ for the O--H stretching mode. Since no gas-phase vibrational spectra of LiOH have been measured, it is not possible to confirm this level of accuracy. However, other molecular \textit{ab initio} PESs constructed in a similar manner have consistently achieved sub-wavenumber accuracy for the fundamentals and many other vibrational levels~\citep{jt612,15OwYuYa1.SiH4,jt652}.

The PES was represented as a tenth-order polynomial expansion of the form,
\begin{equation}
\label{eq:pot}
V =  \sum_{ijk} f_{ijk} \xi_1^{i} \xi_2^{j} \xi_3^{k},
\end{equation}
with the vibrational coordinates,
\begin{eqnarray}
\label{eq:coords_pes}
  \xi_1 &=& (r_1-r_1^{\rm eq})/r_1, \\
  \xi_2 &=& (r_2-r_2^{\rm eq})/r_2, \\
  \xi_3 &=& \alpha-\alpha_{\rm eq}.
\end{eqnarray}
Here, the stretching coordinates $r_1  = r_{\rm LiO} $, $r_2  = r_{\rm OH} $, the interbond angle $\alpha = \angle({\rm LiOH})$, the equilibrium parameters are $r_1^{\rm eq}$, $r_2^{\rm eq}$ and $\alpha_{\rm eq}$, and $f_{ijk}$ are the expansion parameters (44 in total; see Table IV of \citet{13Koputx.LiOH} for values). The PES of LiOH is provided as supplementary material.

\subsection{Dipole moment surface}

A new dipole moment surface (DMS) of LiOH in the \X\ ground state was computed using the coupled cluster method CCSD(T) in conjunction with the correlation-consistent basis sets aug-cc-pCVQZ for Li~\citep{11PaWoPe.ai} and O~\citep{95WoDuxx.ai}, and aug-cc-pVQZ for H~\citep{92KeDuHa.ai}. Calculations used the quantum chemical program MOLPRO2015~\citep{MOLPRO,Molpro:JCP:2020} and were carried out on the same grid of nuclear geometries used to compute the PES, namely 274 points in the range $1.20\leq r_{\rm LiO}\leq 2.40$ and $0.75\leq r_{\rm OH}\leq 1.30$ for the stretches, and $60.0\leq \alpha_{\rm LiOH}\leq 180.0^{\circ}$ for the bending motion.

The instantaneous dipole moment vector $\bm{\mu}$ was represented in the $pq$ axis system~\citep{93JoJexx.H2O}. In this representation, the origin is fixed at the O nucleus with the $p$ and $q$ axes defined in the plane of the three nuclei. The $q$ axis bisects the interbond angle $\alpha = \angle({\rm LiOH})$, while the $p$ axis lies perpendicular to the $q$ axis, e.g.\ at linearity the $p$ axis is along the molecular bond with the Li nucleus in the positive direction. In DMS calculations, the dipole moment components $\mu_p$ and $\mu_q$ were determined via central finite differences by applying an external electric field with components $\pm0.005$~a.u. along the $p$ and $q$ axes, respectively.

Once computed, the $\mu_p$ and $\mu_q$ components were represented analytically using the expressions,
\begin{equation}
\label{eq:mu_p}
\mu_p =  \sum_{ijk} F^{(p)}_{ijk} \zeta_1^{i} \zeta_2^{j} \zeta_3^{k} ,
\end{equation}
and
\begin{equation}
\label{eq:mu_q}
\mu_q =  \sin(\pi-\alpha)\sum_{ijk} F^{(q)}_{ijk} \zeta_1^{i} \zeta_2^{j} \zeta_3^{k} .
\end{equation}
Here, the vibrational coordinates are
\begin{eqnarray}
\label{eq:coords_dms}
  \zeta_1 &=& r_1-r^{\rm eq}_1, \\
  \zeta_2 &=& r_2-r^{\rm eq}_2, \\
  \zeta_3 &=& \cos\alpha-\cos\alpha_{\rm eq},
\end{eqnarray}
with values of $r^{\rm eq}_1=1.6000$~\AA, $r^{\rm eq}_2=0.9478$~\AA, and $\alpha_{\rm eq}=180.0^{\circ}$. A sixth-order expansion ($i+j+k=6$) was used to determine the expansion parameters $F^{(p/q)}_{ijk}$ in a least-squares fitting to the \textit{ab initio} data utilising Watson's robust fitting scheme~\citep{03Watson.methods}. The $\mu_p$ component was fitted with 72 parameters (excluding equilibrium parameters) and reproduced the \textit{ab initio} data with a root-mean-square (rms) error of $1.2\times 10^{-4}$~Debye. For the $\mu_q$ component, 64 parameters achieved an rms error of $6.5\times 10^{-6}$~Debye in the fitting. The DMS of LiOH is provided as supplementary material.

\subsection{Line list calculations}

Line list calculations employed the computer program \EVEREST~\citep{EVEREST}, which was previously used to generate an ExoMol rotation-vibration-electronic line list of the linear triatomic molecule CaOH~\citep{jt878}. We provide only a summary of the key calculation parameters and steps as a full description of the methodology used by \EVEREST\ can be found in \citet{EVEREST}.

Valence bond length-bond angle coordinates were employed with a discrete variable representation (DVR) basis consisting of 100 Sinc-DVR functions on both the Li--O bond in the 2.0--7.0~$a_0$ interval and the O--H bond in the 1.1--6.0~$a_0$ interval, and 120 Legendre functions for the $\angle({\rm LiOH})$ bond angle in the range 0.0 to 180.0 degrees. Vibrational eigenfunctions up to 10\,000~cm$^{-1}$ above the \X\ ground state were computed from a Hamiltonian of dimension 10\,000 for $0\leq K\leq 20$, where $K=|\Lambda + l|$ ($\Lambda$ and $l$ are the projections of the electronic and vibrational angular momenta along the linear axis). The full rotation-vibration (rovibrational) Hamiltonian was diagonalized using the vibrational eigenfunctions for values of $J$ up to 95, where $J$ is the total angular momentum quantum number. Convergence of the computed rovibrational energy levels was tested by increasing the Hamiltonian dimension, setting $K\leq 30$, extending the DVR grids of the coordinates and increasing the number of basis functions. Atomic mass values of 7.014357696863~Da (Li), 15.990525980297~Da (O), and 1.007276452321~Da (H) were used~\citep{AME_2012}. The zero-point energy of LiOH was computed to be 2780.676~cm$^{-1}$.

Rovibrational energy levels, transitions and Einstein $A$ coefficients were computed for the \X\ ground state covering the 0\,--\,20\,000~cm$^{-1}$ range (wavelengths $\lambda > 1$~$\mu$m). The final LiOH line list contained 331,274,717 transitions between 203,762 states with total angular momentum up to $J=95$.

\section{Results}
\label{sec:results}

\subsection{Line list format}

As standard, the ExoMol data format~\citep{jt810} was used to represent the line list as two file types. The \texttt{.states} file, an example of which is given in Table~\ref{tab:states}, lists all the computed rovibrational energy levels (in cm$^{-1}$), a value for the uncertainty (in cm$^{-1}$), and quantum numbers from the \EVEREST\ calculations to identify the states. The \texttt{.trans} file, as seen in Table~\ref{tab:trans}, lists the computed transitions via upper and lower state ID running numbers, and Einstein $A$ coefficients (in s$^{-1}$).

The uncertainties of the energy levels have been estimated as 2~cm$^{-1}$ for the fundamental vibrational states. For excited vibrational states, the uncertainty has been cumulatively added from the contributions of the fundamentals, e.g.\ the state $4\nu_1+\nu_2$ has an uncertainty of 10~cm$^{-1}$. These estimates are conservative and we expect the energies to be more accurate, however, we want to ensure that all users are aware that the LiOH line list is based on theoretical predictions and has not been refined to experimental data. A number of line lists in the ExoMol database have been, or are currently being, adapted for high-resolution applications, namely the high-resolution spectroscopy of exoplanets~\citep{14Snellen,18Birkby}. This is only possible when there is extensive laboratory measurements to replace the computed energy levels by empirically-derived values using the MARVEL procedure~\citep{jt412,07CsCzFu.method,12FuCsxx.methods,jt750}, which is not possible for LiOH due to the lack of experimental data.

The vibrational quantum number labelling of the energy levels was determined in \EVEREST\ by assigning values based on the underlying basis functions used in the variational nuclear motion calculations. The assignments are usually very reliable at lower energies and excitation but become more approximate for highly excited vibrational states, where mixing between the different molecular motions comes into effect. The vibrational quantum numbers used are $v_1$ (symmetric stretching $\nu_1$ mode vibrational quantum number), $v_2$ (bending $\nu_2$ mode vibrational quantum number), $l_2$ vibrational angular momentum quantum number associated with the $\nu_2$ mode, and $v_3$ (antisymmetric stretching $\nu_3$ mode vibrational quantum number).

\begin{table*}
\centering
\caption{\label{tab:states} Extract from the \texttt{.states} file of the LiOH OYT7 line list.}
{\setlength{\tabcolsep}{5pt}
\small\tt
\begin{tabular}{rrrrrrcrcrrr}
\toprule\toprule
$i$ & \multicolumn{1}{c}{$\tilde{E}$} (cm$^{-1}$) & $g_i$ & $J$ & \multicolumn{1}{c}{unc} & $\tau$ & $e/f$ & \multicolumn{1}{c}{State} & $v_1$ & $v_2$ & $L_2$ & $v_3$\\
 \midrule
1 & 0.000000 & 8 & 0 &    0.000000 &    1 & e & X(1SIGMA+) &    0 &  0 &  0 &  0\\
2 & 610.710042 & 8 & 0 &    4.000000 &    1 & e & X(1SIGMA+) &  0 &  2 &  0 &  0\\
3 & 925.369014 & 8 & 0 &    2.000000 &    1 & e & X(1SIGMA+) &  1 &  0 &  0 &  0\\
4 & 1214.983732 & 8 & 0 &    8.000000 &    1 & e & X(1SIGMA+) &    0 &  4 &  0 &  0\\
5 & 1514.667551 & 8 & 0 &    6.000000 &    1 & e & X(1SIGMA+) &    1 &  2 &  0 &  0\\
\bottomrule\bottomrule
\end{tabular}}
\mbox{}\\
{\flushleft
\begin{tabular}{ll}
\toprule\toprule
$i$:  &  State counting number.     \\
$\tilde{E}$: &  State energy (in cm$^{-1}$). \\
$g_i$: &  Total statistical weight, equal to ${g_{\rm ns}(2J + 1)}$.     \\
$J$: &  Total angular momentum.\\
unc: &  Uncertainty (in cm$^{-1}$).\\
$\tau$: &    Total parity. \\
$e/f$:  & Rotationless parity. \\
State: &  Electronic state $X^1\Sigma^+$.\\
$v_1$:  &  Symmetric stretching $\nu_1$ mode vibrational quantum number.\\
$v_2$: &   Bending $\nu_2$ mode vibrational quantum number.\\
$L_2$: &   Vibrational angular momentum quantum number $L_2=|l_2|$ associated with $\nu_2$ mode.\\
$v_3$: &   Antisymmetric stretching $\nu_3$ mode vibrational quantum number.\\
\bottomrule
\end{tabular}
}
\end{table*}

\begin{table}
\centering
\caption{\label{tab:trans}Extract from the \texttt{.trans} file of the LiOH OYT7 line list.}
\tt
\centering
\begin{tabular}{rrrr}
\toprule\toprule
\multicolumn{1}{c}{$f$}	&	\multicolumn{1}{c}{$i$}	& \multicolumn{1}{c}{$A_{fi}$}\\
\midrule
150641 & 153947 & 4.49584894E-02\\
169519 & 168283 & 1.91654138E-01\\
192086 & 190852 & 4.31628948E-02\\
3950 & 5918 & 4.14730097E-09\\
61456 &  62578 & 7.30443603E-06\\
\bottomrule\bottomrule
\end{tabular} \\ \vspace{2mm}
\rm
\noindent
$f$: Upper  state counting number;\\
$i$:  Lower  state counting number; \\
$A_{fi}$:  Einstein-$A$ coefficient (in s$^{-1}$).\\
\end{table}

\subsection{ExoMolOP opacities of LiOH}
\label{sec:opac}

Temperature- and pressure-dependent opacities of LiOH based on the OYT7 line list have been generated for four exoplanet atmosphere retrieval codes ARCiS~\citep{ARCiS}, TauREx~\citep{TauRex3}, NEMESIS~\citep{NEMESIS} and petitRADTRANS~\citep{19MoWaBo.petitRADTRANS} using the ExoMolOP procedure \citep{jt801}. The ExoMolOP temperature grid consisted of 27 values [100, 200, 300, 400, 500, 600, 700, 800, 900, 1000, 1100, 1200, 1300, 1400, 1500, 1600, 1700, 1800, 1900, 2000, 2200, 2400, 2600, 2800, 3000, 3200, 3400]~K; the pressure grid used was $[  1\times 10^{-5},   2.15443469\times 10^{-5},   4.64158883\times 10^{-5},   1\times 10^{-4}, 2.15443469\times 10^{-4},   4.64158883\times 10^{-4},   1\times 10^{-3},   2.15443469\times 10^{-3},
4.64158883\times 10^{-3},   1\times 10^{-2},   2.15443469\times 10^{-2},   4.64158883\times 10^{-2}, 1\times 10^{-1},\times 10^{-1},   4.64158883\times 10^{-1},   1, 2.15443469,   4.64158883,   10,   21.5443469, 46.4158883,   100]$ bar. For the line broadening,  we assumed a 85\% H$_2$ and 15\% He atmosphere  and Voigt line profile with the following parameters: $\gamma_{{\rm H}_2}$ = 0.12~cm$^{-1}$, $n_{{\rm H}_2}$  = 0.5, $\gamma_{{\rm He}}$ = 0.05~cm$^{-1}$ and $n_{{\rm He}}$ = 0.5.

\subsection{Temperature-dependent partition function}
\label{sec:pfn}

The temperature-dependent partition function $Q(T)$ of LiOH has been calculated on a $1$~K grid in the 1\,--\,4000~K range and is provided alongside the LiOH line list from the ExoMol website. Values were obtained by summing over all computed rovibrational energy levels using the expression,
\begin{equation}
\label{eq:pfn}
Q(T)=\sum_{i} g_i \exp\left(\frac{-E_i}{kT}\right) ,
\end{equation}
where $g_i=g_{\rm ns}(2J_i+1)$ is the degeneracy of a state $i$ with energy $E_i$ and total angular momentum quantum number $J_i$, and the nuclear spin statistical weight $g_{\rm ns}=8$ for LiOH.

\begin{figure}
\centering
\includegraphics[width=0.48\textwidth]{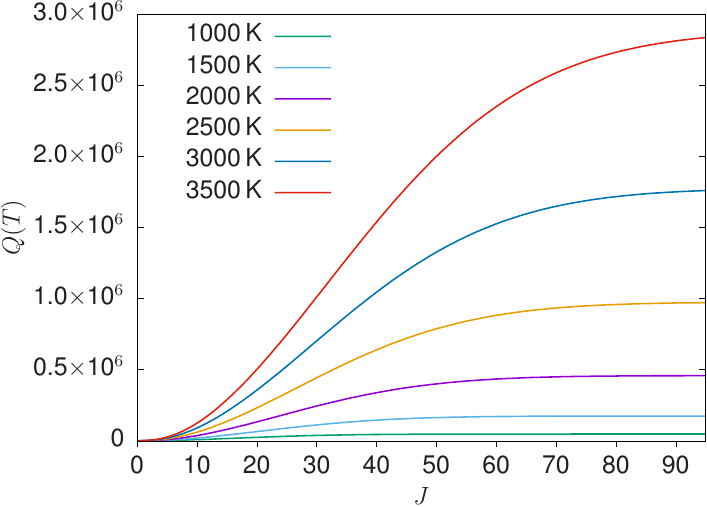}
\caption{\label{fig:pfn}Convergence of the temperature-dependent partition function $Q(T)$ of LiOH with respect to the total angular momentum quantum number $J$ at various temperatures (in Kelvin).}
\end{figure}

In Fig.~\ref{fig:pfn}, the convergence of $Q(T)$ as a function of $J$ is illustrated for different temperatures, essentially showing how more energy levels have to be included in the summation at higher temperatures to achieve converged values. The curve for $T=3500$~K at higher $J$ does not plateau, meaning $Q(T)$ is not fully converged, and we therefore recommend this temperature as a soft upper limit for using the LiOH line list. Using the line list above this temperature will lead to a progressive loss of opacity so caution must be exercised if doing so.

At $T=300$~K, we compute the partition function of LiOH to be $Q=2360.22$. This is much larger than the value available from the Cologne Database for Molecular Spectroscopy (CDMS)~\citep{CDMS:2001,CDMS:2005}, which gives $Q_{\rm CDMS}=1420.69$ (including the contribution from $g_{\rm ns}=8$). This discrepancy can be attributed to the fact that the value from CDMS only considers the ground vibrational state. However, in LiOH there are low-lying rovibrational states that should be properly accounted for. To check this, a partition function value of $Q=1427.80$ was determined from the OYT7 line list by only including contributions from the ground vibrational state in the summation of Eq.~\eqref{eq:pfn}. This comparison highlights the importance of treating rovibrational states in the calculation of the temperature-dependent partition function.

\subsection{Simulated spectra}

\begin{figure}
\centering
\includegraphics[width=0.7\textwidth]{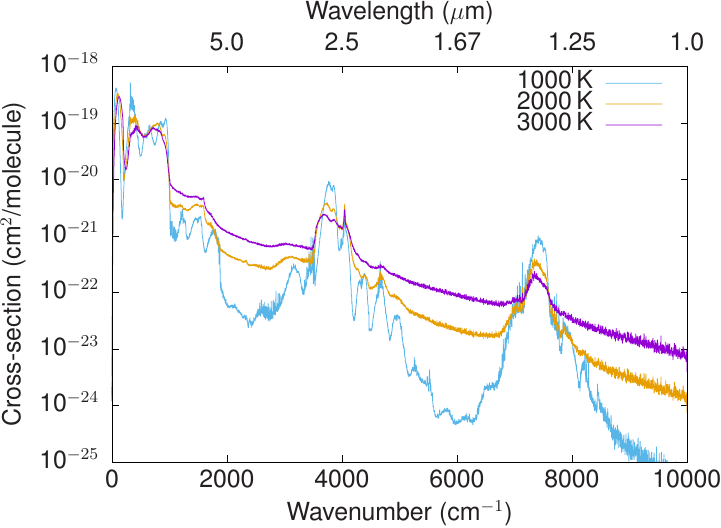}
\caption{\label{fig:1000K-3000K_lioh}Temperature-dependent spectra of LiOH. Absorption cross-sections were computed at a resolution of 1~cm$^{-1}$ and modelled with a Gaussian line profile with a half width at half maximum (HWHM) of 1~cm$^{-1}$ for high temperatures (in Kelvin).}
\end{figure}

An overview of the spectrum of LiOH is shown in Fig.~\ref{fig:1000K-3000K_lioh}, where we have simulated temperature-dependent absorption cross sections at high temperatures. Cross sections were calculated at a resolution of 1~cm$^{-1}$ and modelled with a Gaussian line profile with a half width at half maximum (HWHM) of 1~cm$^{-1}$. Calculations used the \textsc{ExoCross} program~\citep{jt708}, which is based on the methodology presented in \citet{jt542}. As evident in Fig.~\ref{fig:1000K-3000K_lioh}, increasing the temperature causes the rotational bands to broaden substantially, a result of the increased population in vibrationally excited states, leading to much flatter and smoother spectral features. Note that zero-pressure cross-sections of LiOH can be obtained using the ExoMol cross-sections app at \url{www.exomol.com} for any temperature between 100~K and 5000~K, see \citet{jt542}.

\begin{figure}
\centering
\includegraphics[width=0.49\textwidth]{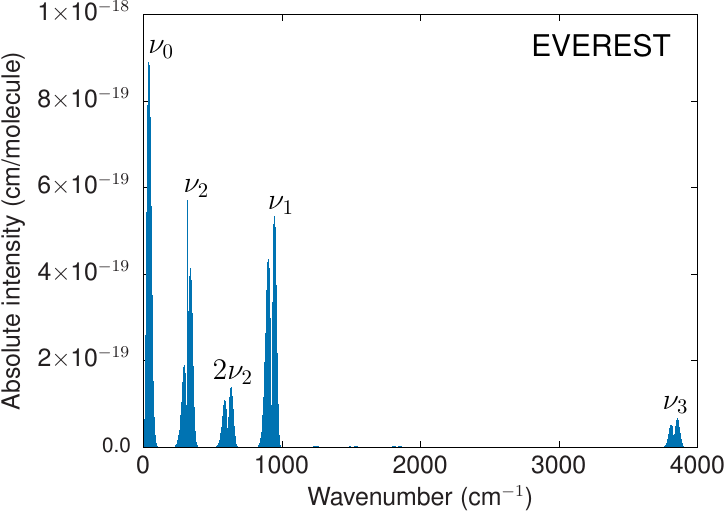}\\
\includegraphics[width=0.49\textwidth]{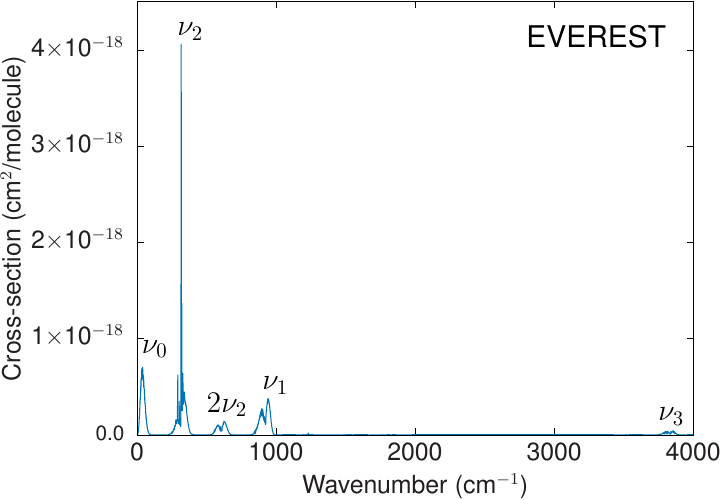}
\caption{\label{fig:stick_lioh}Comparison of computed absolute line intensities (upper panel) and absorption cross sections (lower panel) of LiOH at $T=300$~K. The pure rotational band $(\nu_0)$, Li--O stretching band $(\nu_1)$, Li--O--H bending band $(\nu_2)$, O--H stretching band $(\nu_3)$, and the bending overtone band $(2\nu_2)$ are labelled.}
\end{figure}

The strongest features occur at longer wavelengths where the rotational and fundamental bands lie. In Fig.~\ref{fig:stick_lioh} (upper panel), we have plotted absolute line intensities (in units of cm/molecule) at $T=300$~K showing these bands. Line intensities were computed as,
\begin{equation}
\label{eq:abs_I}
I(f \leftarrow i) = \frac{A_{fi}}{8\pi c}g_{\mathrm{ns}}(2 J_{f}+1)\frac{\exp\left(-E_{i}/kT\right)}{Q(T)\; \nu_{fi}^{2}}\left[1-\exp\left(-\frac{hc\nu_{fi}}{kT}\right)\right] ,
\end{equation}
where $A_{fi}$ is the Einstein $A$ coefficient of a transition with wavenumber $\nu_{fi}$ (in cm$^{-1}$) between an initial state with energy $E_i$ and a final state with rotational quantum number $J_f$. Here, $k$ is the Boltzmann constant, $h$ is the Planck constant, $c$ is the speed of light, $T$ is the absolute temperature, $Q(T)$ is the partition function and the nuclear spin statistical weight $g_{\mathrm{ns}}=8$ for LiOH. The pure rotational band of LiOH appears stronger in intensity than the rovibrationally excited bands when simulating absolute line intensities. However, simulating cross-sections at $T=300$~K (lower panel of Fig.~\ref{fig:stick_lioh}) changes this behaviour and the $\nu_2$ bending mode becomes much stronger than the rotational band, in agreement with previous calculations of the dipole moments of these bands~\citep{04HiFrKl.LiOH}.

\begin{figure}
\centering
\includegraphics[width=0.48\textwidth]{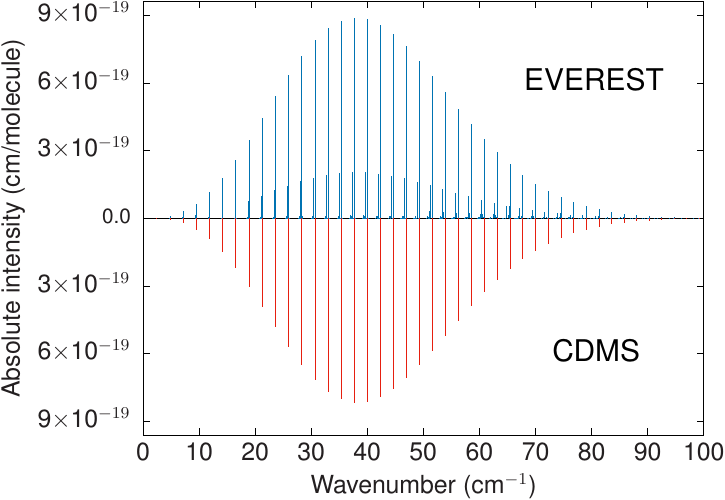}
\caption{\label{fig:rot_lioh}Computed stick spectrum of LiOH at $T=300$~K in the microwave region compared against all transition data from the Cologne Database for Molecular Spectroscopy (CDMS)~\citep{CDMS:2001,CDMS:2005}.}
\end{figure}

The structure of the pure rotational band is shown in Fig.~\ref{fig:rot_lioh}, where we have plotted absolute line intensities at $T=300$~K against experimentally-derived microwave data given by CDMS~\citep{CDMS:2001,CDMS:2005}
based on the measurements of \citet{04HiFrKl.LiOH}. There is good agreement between the strength of our line intensities with the CDMS values computed using a dipole moment value of 4.755~Debye~\citep{04HiFrKl.LiOH}. Overall band shape is reproduced well but our LiOH line list exhibits more structure as we have computed transitions up to $J=95$. The lack of meaningful intensity information on the rovibrational bands of LiOH makes it difficult to quantify the accuracy of our line intensities. Past experience computing \textit{ab initio} DMSs with similar levels of theory suggests that our LiOH transition intensities should be well within 5--10\% of experimentally determined values~\citep{13Yurchenko.method,jt573}.

To encourage future detection of LiOH in exoplanets, we have used the forward modelling capability of the TauREx-III atmospheric modelling and retrieval code to generate example spectra with the rocky super-Earth 55~Cancri~e as the example planet and the system parameters given in \citet{22ItChEd}. We have included LiOH on top of two base scenarios for the atmosphere: a water and carbon dioxide dominated case, and a mineral vapour dominated case.

\begin{figure}
\centering
\includegraphics[width=0.45\textwidth]{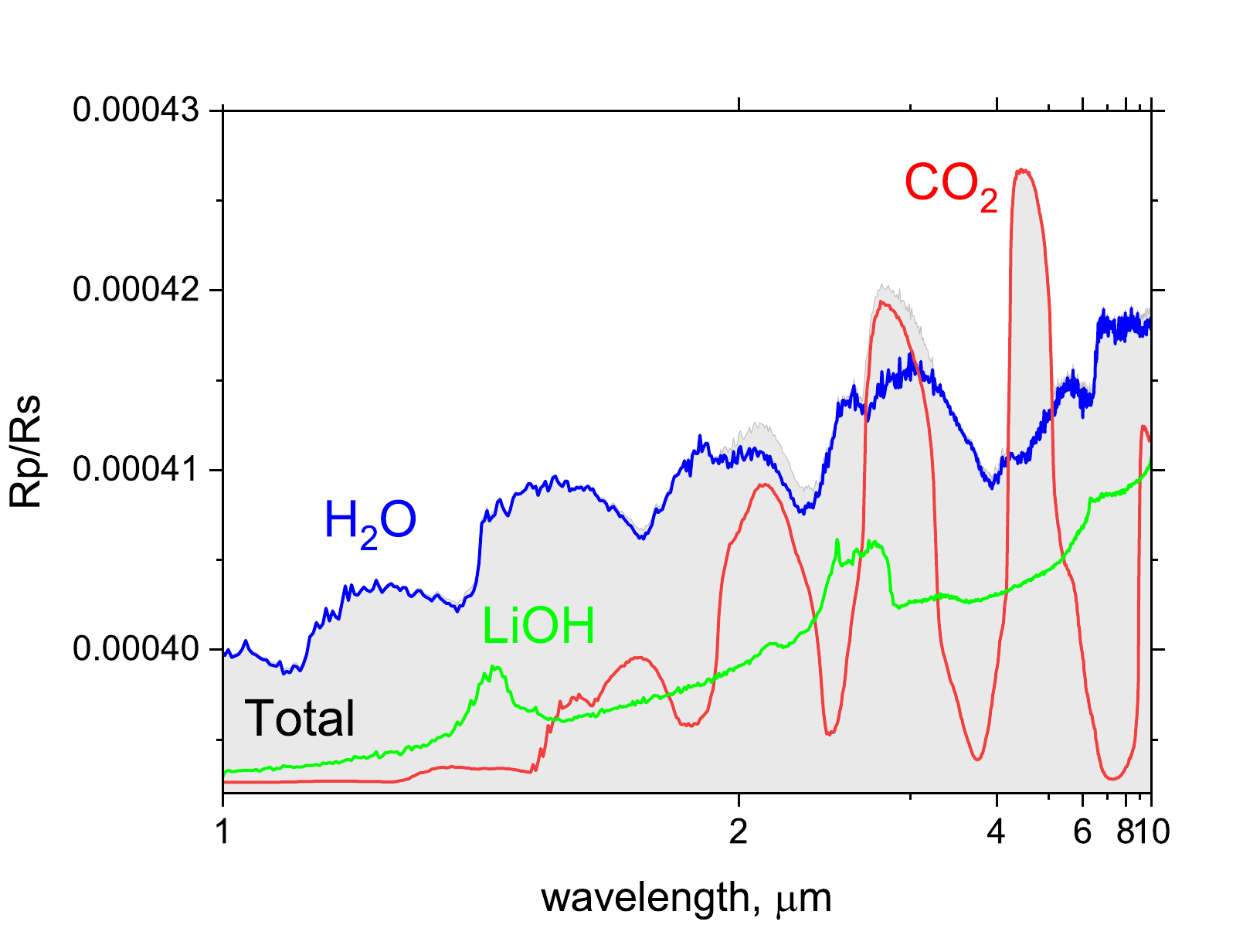}
\includegraphics[width=0.45\textwidth]{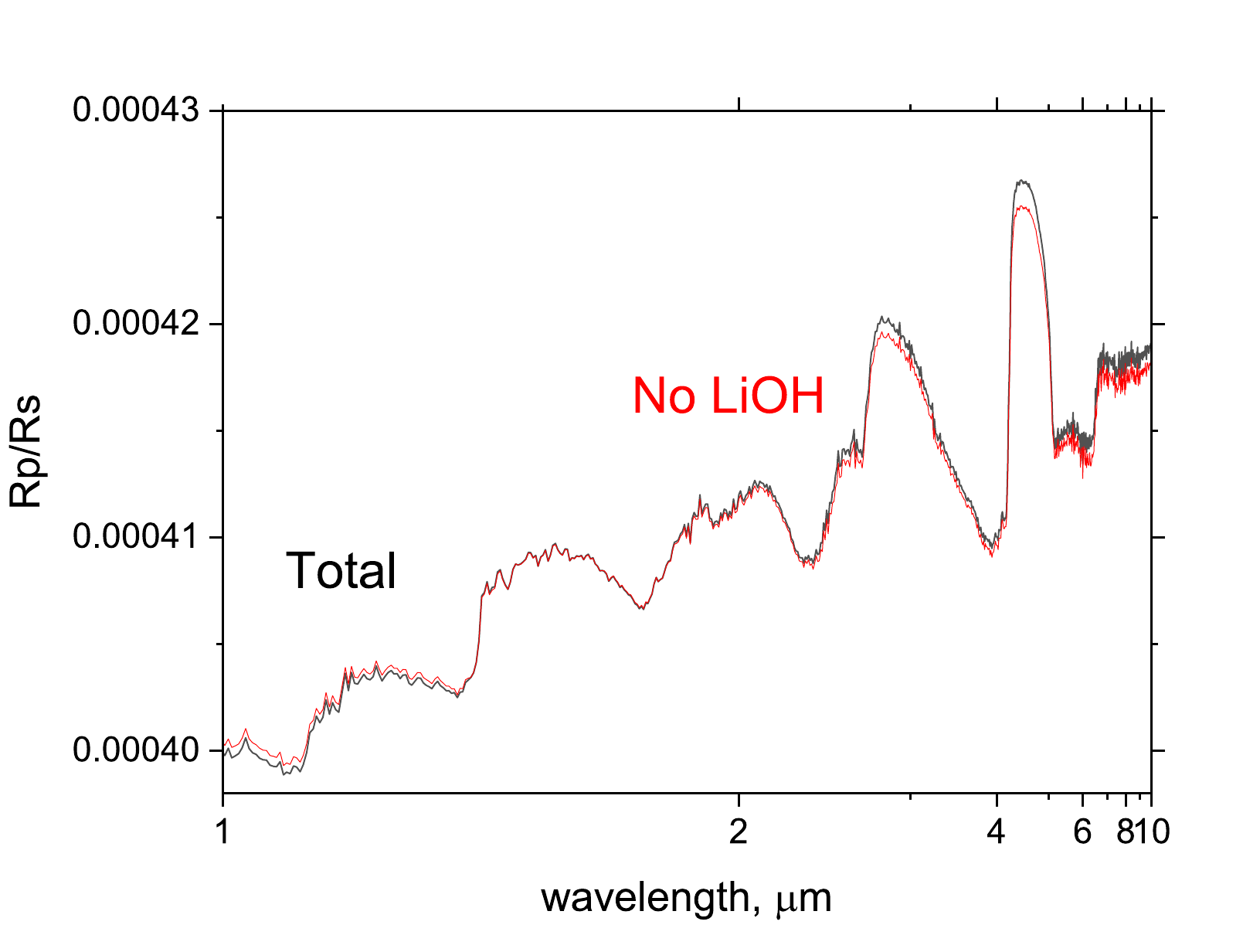}
\caption{\label{fig:transit:H2O} A Simulated transit spectrum of an atmosphere with  45\% of CO$_2$, 45\% of Water and 5\% LiOH using the TauRex-III radiative transfer code \citep{TauRex3} showing how water masks the strongest OH stretching bands of LiOH. The left display shows individual contributions as different colour lines  as well as the total spectrum (grey area). The right display compares the total spectrum (black) with a spectrum where the LiOH contribution was excluded (red) and an offset of $+0.000000796959$ was applied to  latter.}
\end{figure}

The strongest and most promising bands of LiOH in IR (2.6~$\mu$m) and NIR (1.35~$\mu$m) are from the O--H stretching mode, which unfortunately are masked by the H$_2$O bands. As an illustration, Fig.~\ref{fig:transit:H2O} shows a transit spectrum of an (exoplanetary) atmosphere consisting of 50/50 H$_{2}$O, CO$_{2}$ blend diluted down to introduce an example 5\% LiOH contribution, where CO$_2$ also partly overlaps with LiOH's 2.6~$\mu$m band. The NIR (6.7~$\mu$m) band of LiOH is formally in the window, but is probably too weak to make a noticeable difference. The spectra have been produced with the radiative transfer code TauRex-III~\citep{TauRex3} using the OYT7 LiOH opacities generated in this work, while the H$_2$O and CO$_2$ ExoMolOP opacities were produced using the corresponding ExoMol line lists~\citep{jt734,jt804}.

\begin{figure}
\centering
\includegraphics[width=0.47\textwidth]{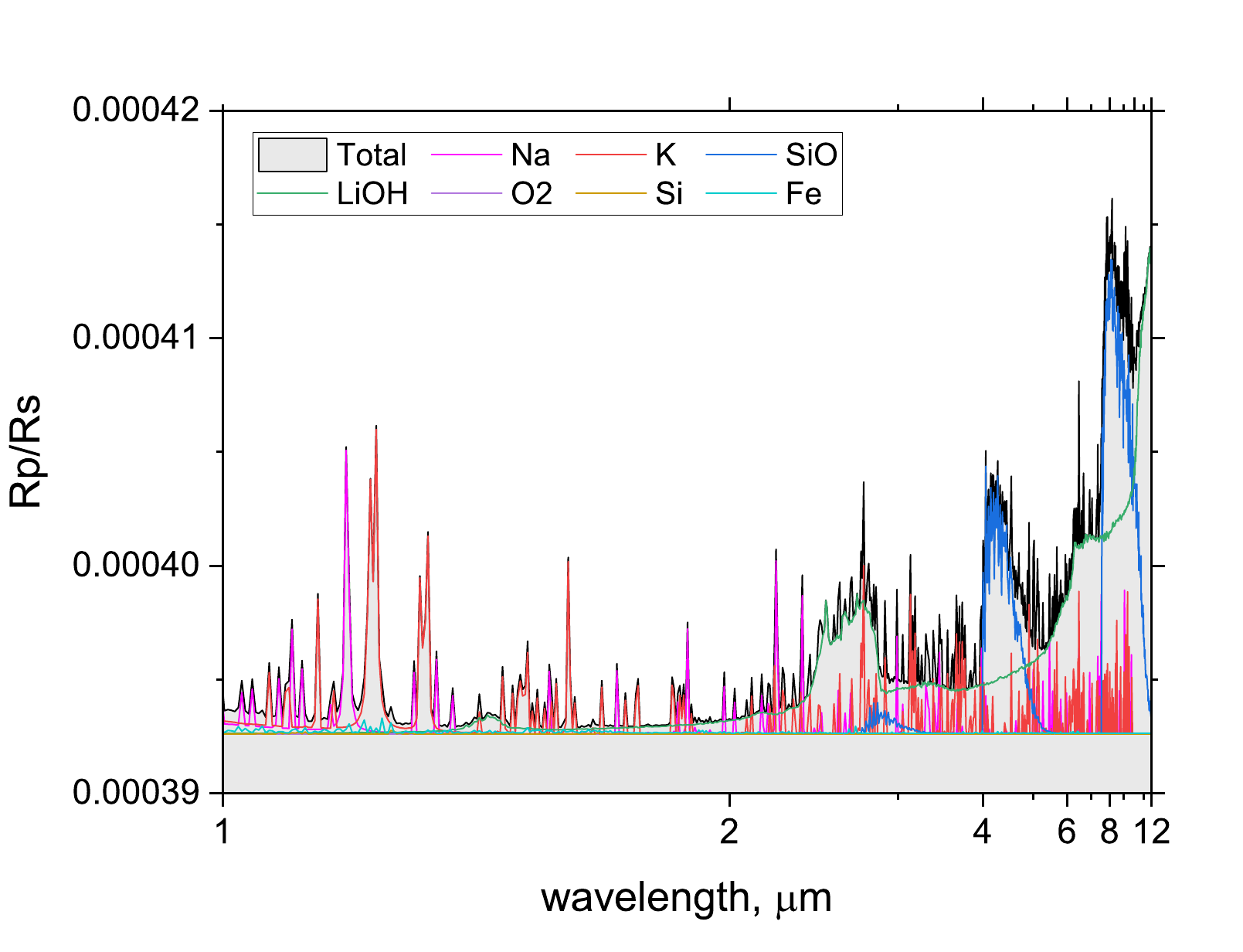}
\includegraphics[width=0.44 \textwidth]{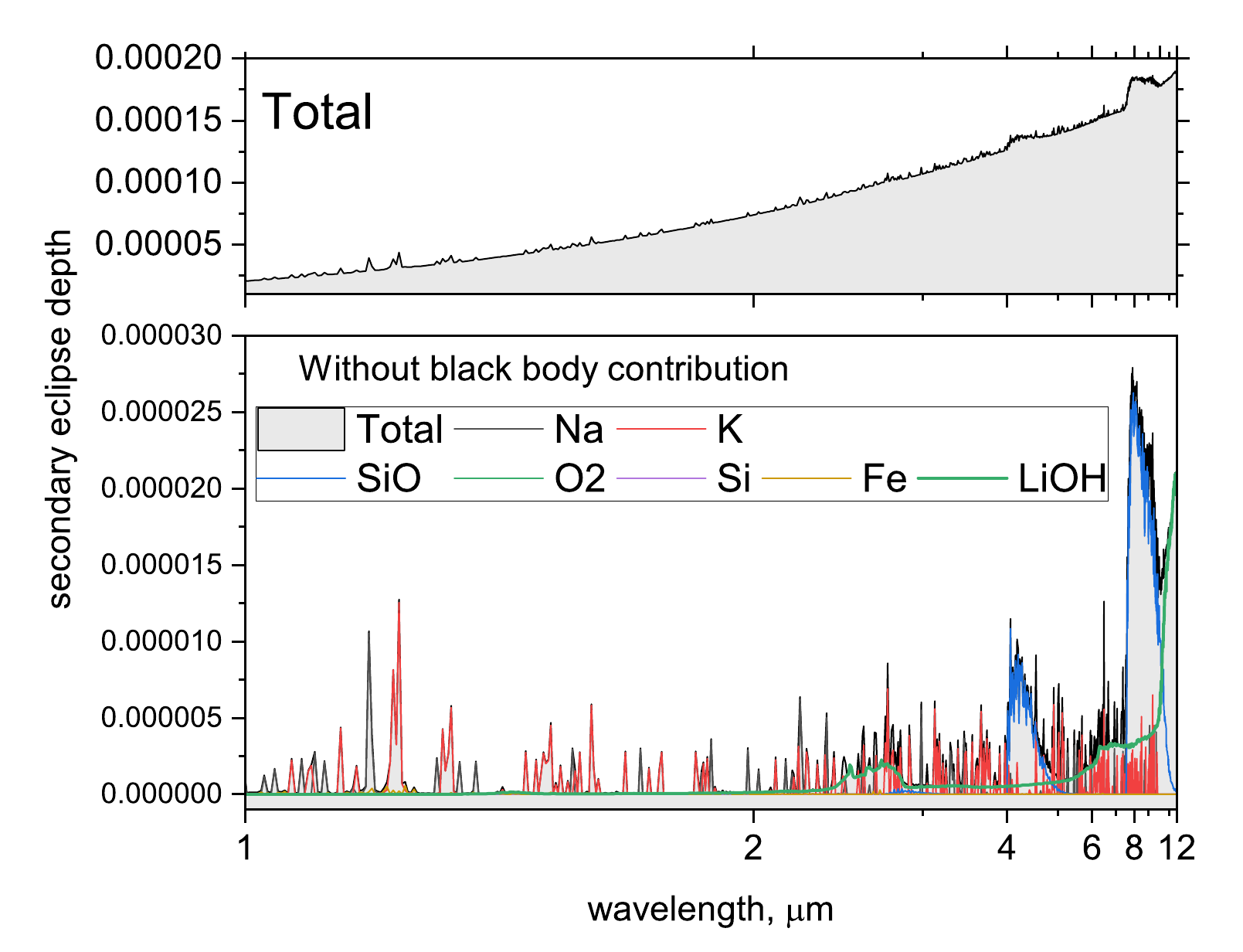}
\caption{\label{fig:mineral} Detectable examples of LiOH in a `mineral' exoplanetary  atmosphere containing Na, K, SiO, O$_2$, Si and Fe, where we added 0.001 part of LiOH (see Fig.~\ref{fig:PT}). Left display: A  transit spectrum (grey background) with all individual contributions shown using different lines with LiOH as a dark green colour. Right display: an emission (secondary eclipse) spectrum of a `mineral' atmosphere (top panel); the lower panel is where the black body radiation background was subtracted and the main individual contributions are shown.}
\end{figure}

\begin{figure}
\centering
\includegraphics[width=0.55\textwidth]{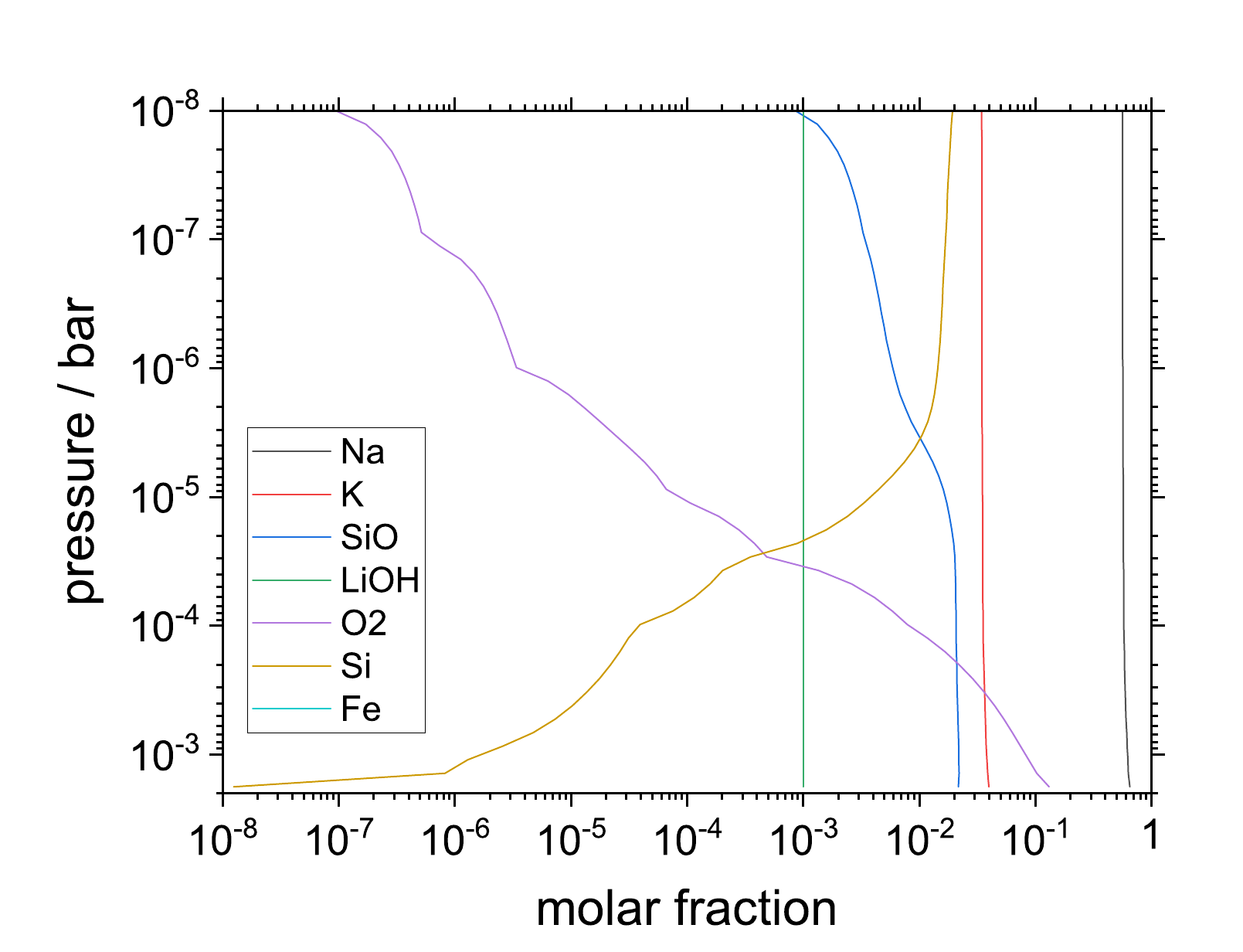}
\caption{\label{fig:PT} Composition of a `mineral' atmosphere as $T/P$ profiles of individual components based on \citet{22ItChEd} with O omitted and LiOH included at an assumed constant value of 0.001.}
\end{figure}

Figure~\ref{fig:mineral} shows a more optimistic scenario of detection, where we have added LiOH to a day-side averaged, mineral-dominated atmosphere from \citet{22ItChEd} containing Na, K, SiO, O$_2$, Si and Fe, with neutral oxygen omitted in favour of an assumed $10^{-3}$ LiOH component (see composition in Fig.~\ref{fig:PT}). The left display gives an example of a transit spectrum, while the right display shows an emission spectrum, where the lower part is with the black body radiation contribution subtracted for clarity.   The underlying line lists are from HITRAN \citep{jt836} for O$_2$, ExoMol for SiO \citep{jt847} and Kurucz for the atomic lines \citep{11Kurucz.db}.

\section{Conclusion}
\label{sec:conc}

A new line list of lithium hydroxide ($^{7}$Li$^{16}$O$^{1}$H) covering wavelengths $\lambda > 1$~$\mu$m (0\,--\,10\,000~cm$^{-1}$) has been presented. The OYT7 line list was computed using the \EVEREST\ nuclear motion code and is based on high-level \textit{ab initio} potential energy and dipole moment surfaces. Transitions and Einstein $A$ coefficients were computed in the \X\ ground electronic state for rovibrational states up to $J=95$. The accuracy of the utilised PES was previously assessed~\citep{13Koputx.LiOH} and we expect line positions of the fundamental bands to be accurate to within 1~cm$^{-1}$ on average, while line intensities should be accurate to within 5--10~\% based on the chosen level of theory used to compute the DMS of LiOH. The strongest IR feature of LiOH corresponds to the $\nu_3$ (O--H stretch) fundamental band at 2.6~$\mu$m which might be difficult to detect in atmospheres of planets and stars due the fundamental band of H$_2$O in this region.

The usual ExoMol methodology is to utilise laboratory data to improve the accuracy of the computed line list~\citep{jt511}. However, the lack of gas-phase spectroscopic measurements of LiOH means that this has not been possible but if data became available in the future then the line list will be updated. A number of purely \textit{ab initio} line lists are available in the ExoMol database and they can greatly aid potential molecular detection. For example, a recent ExoMol \textit{ab initio} line list of silicon dioxide~\citep{jt797} was used to show that SiO$_2$ would be a unique identifier of silicate atmospheres in lava world exoplanets~\citep{22ZiVaMi.SiO2}. It is hoped that the OYT7 line list will assist future astronomical observations of LiOH. As well as the CaOH line list~\citep{jt878} mentioned above, the new OYT7 LiOH line list joins ones for NaOH and KOH already calculated as part of the ExoMol project~\citep{jt820}.

\section*{Acknowledgments}

We thank Yuichi Ito and Quentin Changeat for their help with the exoplanetary  compositions.
This work was supported by the STFC Projects ST/W000504/1. The authors acknowledge the use of the UCL Myriad High Performance Computing Facility and associated support services in the completion of this work, along with the Cambridge Service for Data Driven Discovery (CSD3), part of which is operated by the University of Cambridge Research Computing on behalf of the STFC DiRAC HPC Facility (www.dirac.ac.uk). The DiRAC component of CSD3 was funded by BEIS capital funding via STFC capital grants ST/P002307/1 and ST/R002452/1 and STFC operations grant ST/R00689X/1. DiRAC is part of the National e-Infrastructure. This work was also supported by the European Research Council (ERC) under the European Union’s Horizon 2020 research and innovation programme through Advance Grant number 883830. YP acknowledges financial support from Jes\'{u}s Serra Foundation thought its ``Visiting Researchers Programme'' and from the visitor programme of the Centre of Excellence ``Severo Ochoa'' award to the Instituto de Astrofísica de Canarias (CEX2019-000920-S). SW was supported by the STFC UCL Centre for Doctoral Training in Data Intensive Science (grant number ST/P006736/1).

\section*{Data Availability}

The states, transition, opacity and partition function files for the LiOH line list can be downloaded from \href{https://exomol.com/data/molecules/LiOH/7Li-16O-1H/OYT7/}{www.exomol.com} and the CDS data centre \href{http://cdsarc.u-strasbg.fr}{cdsarc.u-strasbg.fr}. The open access program \textsc{ExoCross} is available from \href{https://github.com/exomol}{github.com/exomol}.

\section*{Supporting Information}
Supplementary data are available at MNRAS online. This includes the potential energy and dipole moment surfaces of LiOH with programs to construct them.


\bsp	
\label{lastpage}
\end{document}